\documentclass[twocolumn,showpacs,preprintnumbers,amsmath,amssymb]
{revtex4-1}
\usepackage{graphicx}
\usepackage{dcolumn}
\usepackage{array}

\graphicspath{%
    {converted_graphics/}
    {/}
}
\usepackage[english]{babel}
\usepackage[latin1]{inputenc}
\begin{document}


\title{The Effect of Varying Co layer thickness on the Time-Temperature Characteristics of Co/Sb Semimetal Embedded Magnetic Nanoparticles}

\author{M.R. Madden, T. Alshammary, B. Grove, J. Phillips, K. Reaz, S. Hensley and G.G.~Kenning.}

\affiliation{Department of Physics, Indiana University of Pennsylvania,
Indiana, Pennsylvania 15705-1098}

\date{\today}

\begin{abstract}
We report the effect  of varying cobalt thickness on the temperature-dependent time decay of the electrical resistance  of Co/Sb multilayer samples.  We find that for a given temperature, a five fold change in the Co thickness produces a 100 fold change in the characteristic decay time of the resistance. We find that the characteristic decay time, as a function of temperature, follows an Arrhenius law. During deposition, the Co evolves single domain magnetic nanoparticles, on the Sb, in either a Volmer-Weber or Stranski-Krastanov growth  mode. This metastable state is then encased in 2.5 nm of Sb producing an embedded nanoparticle system.  Scanning Tunneling Microscopy (STM) measurements taken before sample aging (annealing at a given temperature for enough time to complete the resistance decay) and after aging show that these nanoparticles undergo morphological transformations during aging. These transformations lead to well defined time dependent decays in both the magnetization and the electrical resistance, making this material  an excellent candidate for an electronic time-temperature sensor.

\end{abstract}

\maketitle


\section{Introduction}
\label{sec:SclTemp1}

Over the last 20 years advances in nanotechnology have opened the way to new types of applications and devices. Materials at the nanoscale often display physical properties very different from the bulk material, and these new properties have been harnessed for a variety of  electronic \cite{Wasser12}, photonic \cite{Prasad04}, and medical \cite{Prasad12} applications. This paper is focused on investigations of a new type of nano-electronic device that exploits the time-temperature properties of Co nanoparticles encapsulated within an Sb matrix. In  previous studies \cite{Kenning11}\cite{Kenning14}, we have reported the observation of magnetic and resistive aging in a self assembled nanoparticle (Co) system produced in a Co/Sb multilayer  sandwich.  Both the temperature  and time scales are in potentially useful regimes (near room temperature and minutes to months) and these materials have possible application as time-temperature indicators.  The relationship between the characteristic time and aging temperature fits an Arrhenius law indicating activated dynamics. The Arrhenius law also governs the decay of many types of perishable foods and pharmaceuticals. We have been able to produce materials with time-temperature characteristics similar to that found in meat and milk. \cite{Lu12} Therefore, we see the possibility that this material may be used as an electronic time-temperature indicator (TTI) to monitor the lifetime of these products. Chemical time-temperature indicators have found commercial application as efficacy indicators for pharmaceuticals \cite{Robertson93}. These indicators change color based on chemical reactions following an Arrhenius law that is designed to match the Arrhenius parameters of a particular pharmaceutical. If the pharmaceuticals efficacy has decayed due to time and temperature stress, this is recorded in the indicator's color change. 

An electronic time-temperature indicator coupled with rf sensing, such as RFID sensing, would enable significant added functionality. The ability to automatically analyze the thermal age of products along the cold chain (temperature controlled supply chain) would provide for automated decision making and reduced waste. An inexpensive sensor would enable  monitoring of individual consumer items such as milk, meat, fish, and pharmaceuticals. The Natural Resources Defense Council \cite{Gunders12} estimates that almost 60$\%$ of fish, 23$\%$ of beef and 21$\%$ of drinking milk is wasted. The NRDC has also concluded that the vast majority of this waste is at the level of the consumer. This makes sense from the perspective that the entire cold chain runs off of a ``sell by" date. Thermally stressed food is pushed onto the consumer as long as the ``sell by" date has not been reached. Most cell phones now contain rf readers, which coupled to the above envisioned sensor, would give consumers the ability to query the actual age and projected shelf life of the products they purchase.

In general, time-dependence in electronics has been focused on two main properties: 1) making  components last longer and 2) minimizing the on/off time factors to make components faster. To date only limited work has been done on exploiting the time dependence of a components' lifetime to produce new functionality. 

The study of long time dependencies in out of equilibrium ensembles of chemical constituents goes back to the 1800's and was highlighted with the seminal work of Arrhenius. He derived a time-temperature dependent law for thermally activated transitions in a two-state system, where the states are separated by an energy barrier. Since this time the Arrhenius law has been found to be ubiquitous in nature, describing such diverse phenomena as chemical reactions \cite{Arrhenius1889}, the decay of many types of food due to microbial growth 
\cite{Labuza84}, and spin flipping in superparamagnets \cite{Neel49} to name a few.

Prior to the advent of the PC, precise long time measurements were difficult to make.  In general, materials whose physical characteristics changed over long time scales were avoided as they made reproducibility difficult on practical timescales. One of the first detailed sets of time measurements were made by Struick \cite{Struik78} on the time dependence of the strain in polymers subjected to a constant stress for set amounts of time. Long time dynamic measurements on spin glasses \cite{Cham83} and electron glasses 
\cite{Amir08} have yielded programmable time-dependent memory effects. In all of these systems, the initial state is prepared in a field for a given amount of time $t_1$. This is often called the waiting time $t_w$. The field is then changed and the long time response, often a decay, establishes a characteristic time scale $t_2$ which is dependent on the preparation time $t_1$. These effects are also strongly influenced by temperature changes during the waiting time and during the decay. Unfortunately the temperature range over which these effects exist has so far precluded practical application.

Previous magnetic measurements of Co/Sb multilayer films displayed evidence of superparamagnetism, with a blocking temperature around 100 K, indicating the formation of Co nanoparticles. \cite{Heidt09}  At higher temperatures (300-400 K), it was observed that the magnetization underwent a large decay.  The formation of nanoparticle systems, that decayed, was found to be stongly dependent on the deposition temperature as well as the substrate and the time allowed for the Co to migrate, before the next Sb layer was deposited. In the Co/Sb system, this is the equivalent preparation time $t_1$. In the study presented here, all samples were deposited on a 50 $^oC$ substrate and in general, the 1.0 nm samples took about 60 sec. to deposit. After deposition, the Co was allowed to migrate on the Sb surface for 300 sec..  The total  time $t_1$,  includes both the wait time and a more complicated time scale associated with the deposition time. The Co layer is then encased by deposition of 2.5 nm of Sb, locking in the the time scale $t_1$. The total magnetization decay was as much as 80$\%$ of the initial magnetization and the time-temperature behavior of the decay followed an Arrhenius law. The DC electrical resisitance was also found to decay and had the same time-temperature characteristics (Arrhenius parameters) as the magnetization decays. \cite{Kenning11}  Temperature dependent resistance measurements (R vs T) show that before aging, the samples displayed negative dR/dT indicating semiconductor behavior.  As the sample was aged, dR/dT underwent a transition as a function of aging time, crossing over to positive dR/dT for a fully aged sample, indicating metallic behavior. \cite{Kenning14}

The combination of a magnetic metallic constituent (Co) and a semi-metal (Sb) in an embedded magnetic nanoparticle material produces some very interesting properties.  Sb is a Group V semimetal with an orthorhombic crystal structure. Its semimetallic behavior is based on the projection of the valence band at the H point above the Fermi energy  and the conduction band at the L point extending  below the Fermi energy. \cite{Issi79}\cite{Liu95} The  overlap between these bands is 180 mK at 4.2 K. \cite {Heremans01} Under specific circumstances, semimetals can behave as intrinsic semiconductors, \cite{Halperin68} and semimetal-semiconductor transitions have been observed in Bi doped with In or Sb. \cite{Rani02} Heremans al. have observed a negative dR/dT  in 48 nm diameter Sb nanowires in small magnetic fields. \cite {Heremans01} Co is a transition series magnetic atom that forms a ferromagnetic metal in bulk.  The carrier density  of Co is approximately 2-3 orders of magnitude greater than that of bulk Sb.

In this work, we have extended our time dependent measurements of Co/Sb nanoparticle systems to include variations in the deposited Co layer thickness. The Co nanoparticles self-assemble shortly after deposition, and in this manuscript we describe the Co deposition in terms of the layer thickness deposited. We further analyze STM images of Co/Sb nanoparticle systems of varying Co layer thicknesses. STM imaging was performed both before and after aging the material.

\section{Experimental Methods}
Multilayer samples were prepared in an Edwards E306A e-beam evaporator with a starting pressure of 2x10$^{-7}$ mbar. Samples were deposited on the $<100>$ face of silicon held at a temperature of 50 $^oC$. The deposition rates were continuously monitored with a Sycon thin film deposition rate/thickness monitor located adjacent to the sample holder.  The deposition rate was 0.02 - 0.03 nm/sec.. All samples had Sb layer thicknesses of 2.5 nm. For all samples reported here, an initial 2.5 nm layer of Sb was deposited on the bare substrate and allowed to equilibrate for 200 sec. before deposition of the next layer of Co. After the Co was deposited, the samples were allowed a waiting time of 300 sec. before the next layer of Sb was deposited. This process was repeated until ten bilayers were produced for resistance measurement. Finally the samples were capped with a final 2.5 nm layer of Sb. For microscopy measurements, Sb (2.5 nm)/Co(x)/ Sb (2.5 nm) (x=0.5 nm, 1.0 nm and 5.0 nm) samples were prepared (1.5 bilayers). The deposition system is not automated, requiring manual gun rotation as well as manual deposition temperature control. Sample storage considerations are discussed elsewhere. \cite{Kenning14}  Samples imaged in the STM were measured in two stages.  First, freshly prepared samples were imaged with the STM at room temperature. These images are designated ``Before Aging" in the manuscript. The samples were then removed from the STM and heated at 400 K for 3600 sec..  The sample was removed from the furnace and placed back in the STM for another room temperature scan. These second measurements, are designated ``After Aging". The sample remained on the magnetic sample holder for the entire process.

All resistance measurements in this study were conducted using the  MagnetoResistance (MR) probe option on a LakeShore Model 7307 Vibrating Sample Magnetometer (VSM). After loading the sample onto the MR probe and measuring the room temperature resistance, the sample was inserted into the cryostat, which was preheated to the measuring temperature. All resistance measurements in this study were made with a 4-probe DC technique except the Co/Sb (1.5 nm/ 0.8 nm) sample which was measured with the 2-probe technique. The 4-probe technique is an in-line measurement with the current leads  set 3 mm apart and the voltage leads separated by 1 mm, located in-line 1 mm from the respective current leads. The 2-probe measurement sensed the voltage drop across the current leads. All resistances were measured in zero magnetic field.

STM measurements were made using a Nanosurf easyScan E-STM system. STM tips of PtIr were prepared by cutting the wire at a prescribed angle and electro-etching the tip in 1.5M CaCl$_{2}$ solution. Tip quality was gauged by performing a 4 nm x 4 nm STM scan on highly oriented pyrolytic graphite (HOPG). If atomic resolution on this scale was observed, the tip was deemed suitable for imaging the Co/Sb samples.  The STM was run in constant current mode with the gap voltage set on the 30-50 mV range. In this mode, the STM scans the surface maintaining a constant tunneling current of 1 nA. Variations from 1 nA provide feedback to the z component of the piezoelectric electric tip driver, leading to the tip withdrawing if it senses a larger current or moving toward the sample if it senses a smaller current. For a homogenous sample the STM scan maps the topography of the surface. The tunneling current is, however, not solely dependent on distance to the surface; it depends primarily on the density of energy states available for the tunneling current to occupy at and below the surface. Therefore, inhomogeneities under the surface can be observed if they change the Local Density of States (LDOS). STM images were analyzed with Scanning Probe Image Processor (SPIP) (Image Metrology) ApS.software.

\section{Resistance Decay data as a function of Co thickness}

Fig.~1 displays the normalized resistance decay curves for samples with six different deposited layer thicknesses of Co.  We have normalized the resistance on the y-axis and plot R/R$_{max}$ vs. ln(t) using the same scale for each sample. It can be readily observed that as a function of Co thickness several systematic changes occur. At a Co layer thickness of 0.5 nm, no resistance decay was observed over the temperature range 340 K to 440 K. For the samples with 0.8 and 1.0 nm Co layer thicknesses, a complete set of resistance decays were observed for temperatures ranging from 290 K to 340 K and 330 K to 400 K respectively. For these sets of samples, measurements were taken every 10 K and the systematic shift in the curves on a log scale suggests a temperature resolution of about $\pm$2K. For these samples, the total decay of the resistance is approximately 50$\%$ of R$_{max}$.  As the Co thickness is further increased, we observe that the total decay of the resistance decreases, and at 5.0 nm the total decay is only about 10$\%$ of R$_{max}$.

\begin{figure} 
\centering 
\centerline{\includegraphics[width=3.15in,height=3.68in,keepaspectratio]{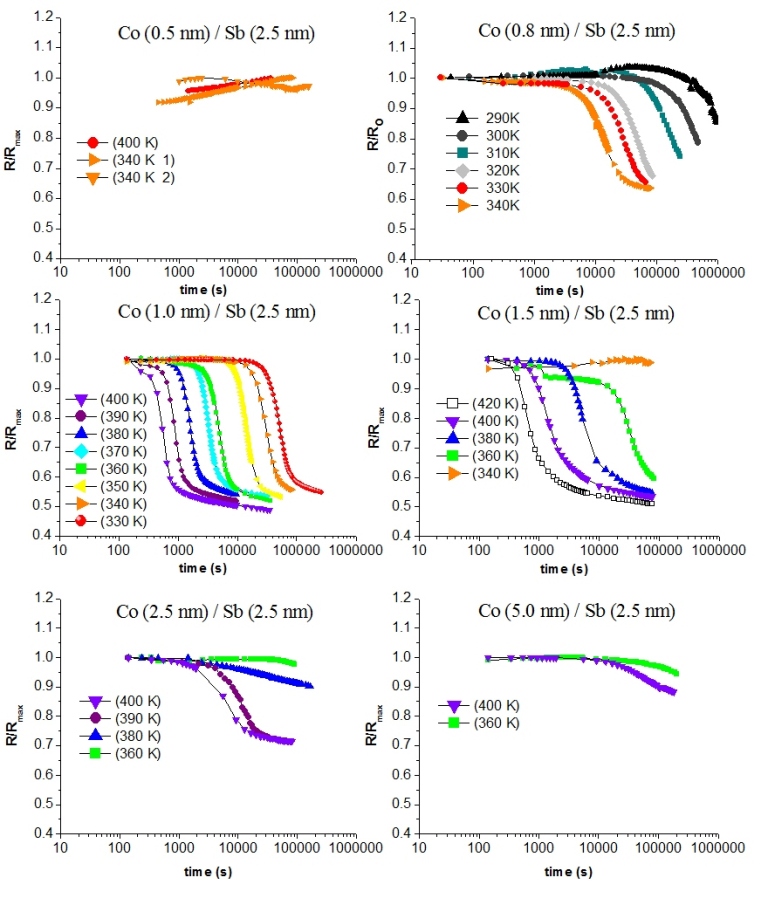}}

\caption{\label{fig:bob9} Normalized resistance decay curves for Co$(x)$/Sb $(2.5~nm)$ with x = 0.5, 0.8*, 1.0, 1.5**,  2.5 and 5.0 nm.  The curves were normalized to the maximum resistance observed in the decay. 
*reproduced from\cite{Kenning14}
**reproduced from\cite{Kenning11}}
\end{figure}

\begin{figure} 
\centering 
\centerline{\includegraphics[width=3.15in,height=2.61in,keepaspectratio]{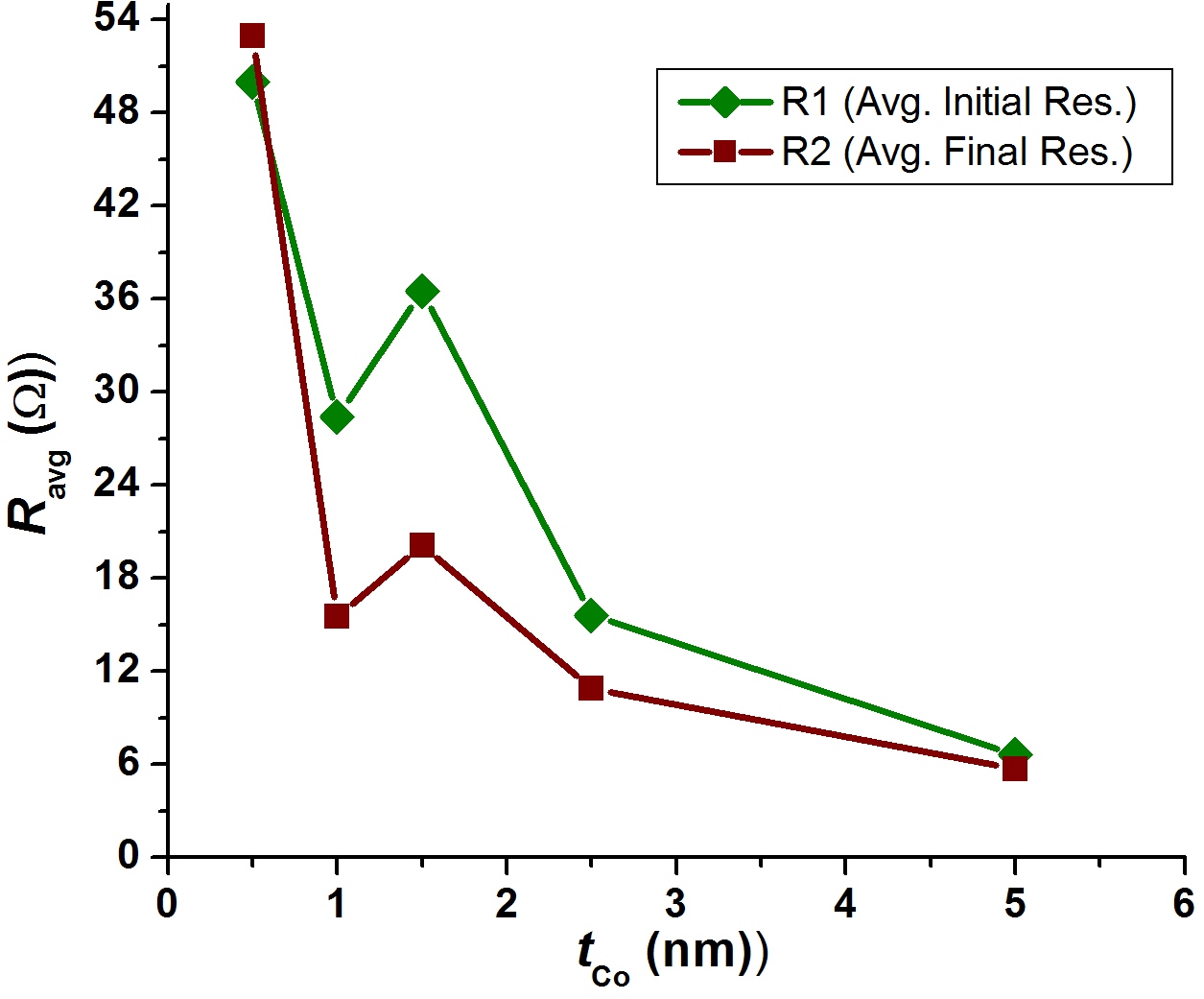}}

\caption{\label{fig:bob8} Average resistance of samples before and after aging, as a function Co layer thickness $t_{co}$.}
\end{figure}

In Fig.~2 we plot the average resistances of the samples used in the study, before and after aging. We use averages as there are  sample to sample resistance variations due to slight differences in the sample sizes. The 0.8 nm sample was measured using the 2-probe technique (resistance values were much larger than the 4-probe technique) and is excluded from this analysis. The region between the before and after aging resistances curves gives a remanent resistance, which is the resistance component that decays.  It can be observed that the resistance difference has a maximum value for Co thicknesses of 1.0 nm and 1.5 nm. The 0.5 nm sample has the largest initial resistance. For the 2.5 nm and 5.0 nm samples, the remanent resistance is small, suggesting that the mechanism responsible for the decay is only a small part of the sample resistance. We also notice that doubling the Co thickness in this regime (2.5 nm to 5.0 nm) reduces the resistance by approximately half. From this we conclude that the Co is effectively continous in the thicker samples and provides an important channel for the flow of electrons through the sample.

The most important systematic change in the data is the time dependence of the decays as a function of Co layer thickness. A cursory observation of the 400 K measurements shows that the decay time increases significantly as the Co thickness increases. This effect will be discussed quantitatively in the next section.

\begin{figure} 
\centering 
\centerline{\includegraphics[width=3.15in,height=2.5in,keepaspectratio]{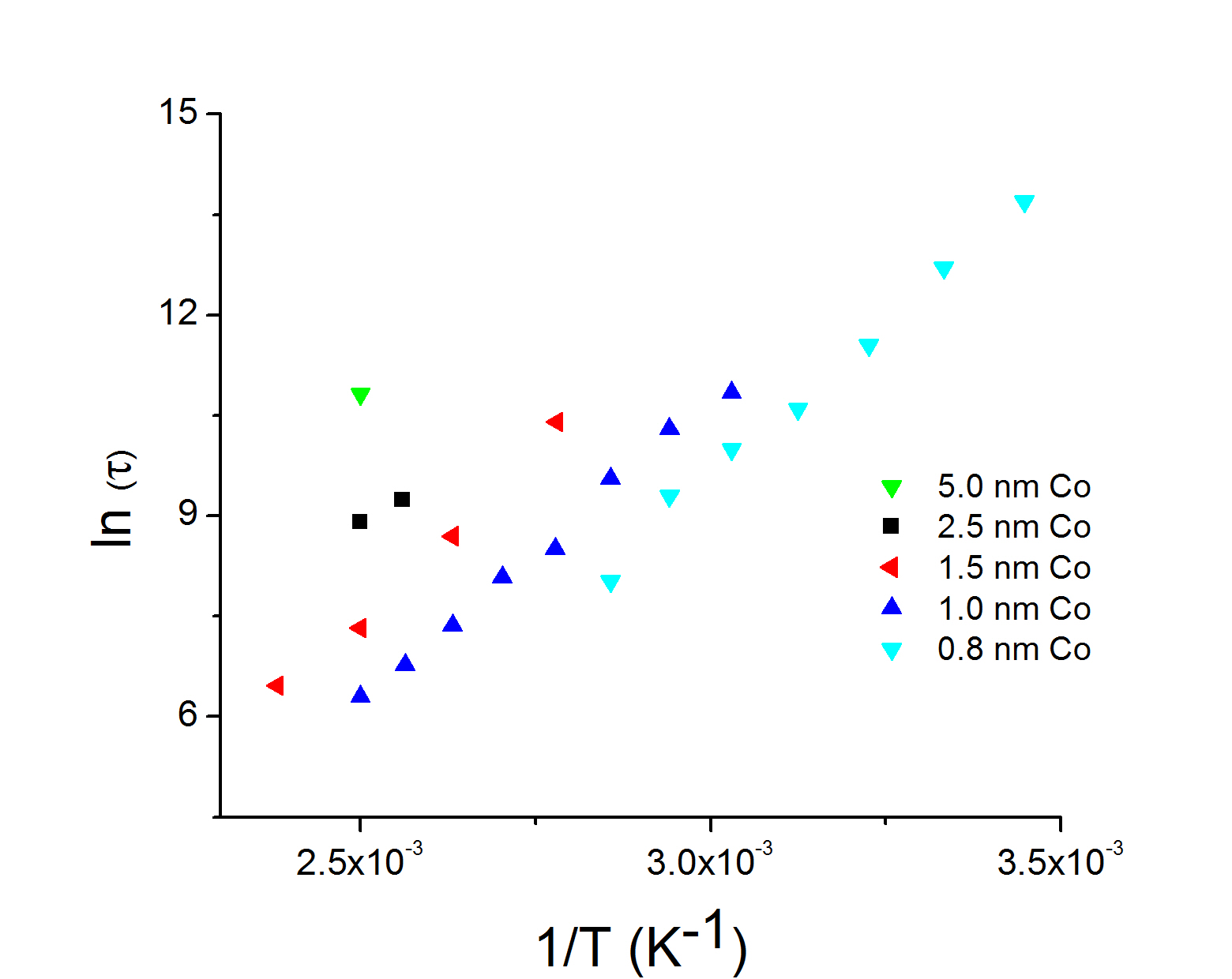}}

\caption{\label{fig:bob8} Arrhenius plots for samples in Fig 1.  }
\end{figure}

\subsection{Arrhenius Analysis as a Function of Co Layer Thickness}

From the data in Fig.~1 we were able to extract a single characteristic time $\tau$ associated with a full decay for each temperature run. The decay time $\tau$ was extracted from an individual curve by first fitting the resistance decay to a sigmoidal function and taking the function's derivative with respect to ln(t). The quantity -dR/d(ln(t)) has a maximum at the inflection point. We designate the time associated with the inflection point as the decays characteristic time $\tau$. The only exception to this technique was the 0.8 nm Co sample (reproduced from ref. \cite{Kenning14}), where the characteristic time  was obtained from a single resistance value 
$R=.85R_0$, where $R_0$ is the initial sample resistance before the decay process begins. In previous studies of Co/Sb samples \cite{Kenning11} we found that the decay time $\tau$ follows a Neel-Arrhenius law as a function of temperature. The data can be analyzed by plotting a straight line to the data on a ln($\tau$) vs. 1/T plot using the equation;

\begin{eqnarray}
\tau=\tau_o exp(E_a/RT),
\end{eqnarray}

From this analysis, the slope of the line determines the activation energy and the y-intercept provides the fluctuation timescale $\tau_0$. Table 1 displays the values for the activation energy and fluctuation time scale $\tau_o$. 

\begin{table}[t]
\caption{Values of the activation energy $E_a$ and the fluctuation time scale $\tau_o$ for three Co/Sb samples. \cite{correction} }

\begin{tabular}{ccc}    
\hline    
\hline
~~~~~~~Sample~~~~~~~~~~~~~~ & ~~~$E_a (kJ/mol)$~~~~~~~~~~~~~~~~ & $\tau_o (s)$\\    
\hline

Co/Sb (1.5 nm/2.5 nm) &  82~~~~~~~~~~& 2.7x10$^{-8}$\\
\\
Co/Sb (1.0 nm/2.5 nm) &  74~~~~~~~~~~& 1.1x10$^{-7}$\\
\\
Co/Sb (0.8 nm/2.5 nm) &  77~~~~~~~~~~& 1.4x10$^{-8}$\\

\hline
\hline
\end{tabular}
\\

\label{tab:multicol}
\end{table}

It is clear from Fig.~3 that for a given measuring temperature, the characteristic time  shifts to larger values for larger Co thicknesses. For example, at a measuring temperature of 400 K (${1 \over T} =2.5x10^{-3}~K^{-1}$), $\tau$ shifts systematically from $\approx$400 sec. for a Co layer thickness of 1.0 nm, to $\approx$1000 sec for a 1.5 nm Co layer.   Increasing the Co layer thickness to 2.5 nm increases $\tau$ to $\approx$7000 sec.  and finally a 5.0 nm Co layer thickness has a $\tau$ of $\approx$40,000 sec.. In other words, a factor of five change in the layer thickness causes a change of a factor of 100  in the decay time $\tau$. The timescale is therefore highly tunable through the variation of  Co thickness. Tunability is an important aspect necessary for the application of this material to a wide variety of time-temperature indicators. Different foods or pharmaceuticals have distinct Arrhenius behaviors, and in each case a time-temperature indicator will need to be tuned to the specific application.

\begin{figure}[t]
\centering 
\textbf{Sb/Co/Sb (2.5 nm/0.5 nm/2.5 nm)}\par\medskip
\textbf{Before Aging}\par\medskip
\centerline{\includegraphics[width=3.15in,height=2.8in,keepaspectratio]{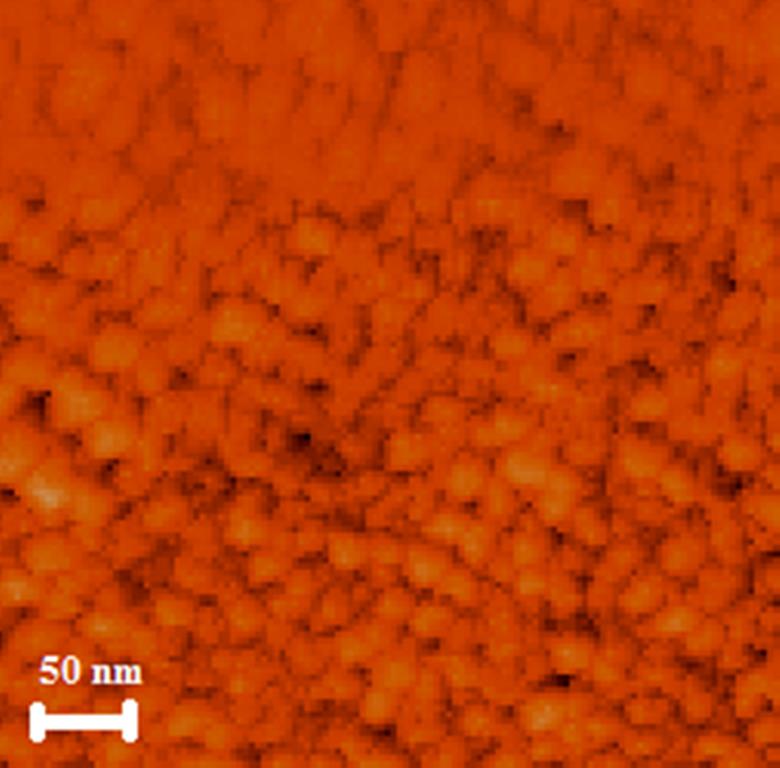}}
\caption{\label{fig:bob1} Scanning Tunneling Microscopy (STM) image of the Sb/Co/Sb (2.5 nm/0.5 nm/2.5 nm) sample before aging. \medskip}
\end{figure}

\section{Scanning Tunneling Microscopy (STM)}
	
STM was performed on three different samples (0.5 nm, 1.0 nm and 5.0 nm Co layer thickness), before aging at a constant temperature, and then after aging. Unfortunately, due to a lack of a heater on the STM, the sample had to be removed from the STM, aged and then placed back in the STM for the after aging scan. With this protocol it was not possible to observe the exact same area of the sample. However, due to sample alignment within the STM the two regions are fairly close together, within a few hundred microns.

Images were analyzed with SPIP (Image Metrology) ApS.software using a threshold detection method for particle analysis. The software numerically performed two independent calculations to determine the particle's area and perimeter. Once the area calculation is complete, the program then assumes  the particle to be circular  and  calculates the diameter. The ratio $r = perimeter/diameter
= \pi$ for circular particles. Deviations from $\pi$ will be analyzed.   The images presented in this manuscript have been contrasted and enhanced using Microsoft Picture Manager and/or Adobe Photoshop.

\begin{figure}[t] 
\centering 
  \textbf{Sb/Co/Sb (2.5 nm/0.5 nm/2.5 nm)}\par\medskip
 \textbf{After Aging}\par\medskip
\centerline{\includegraphics[width=3.15in,height=2.8in,keepaspectratio]{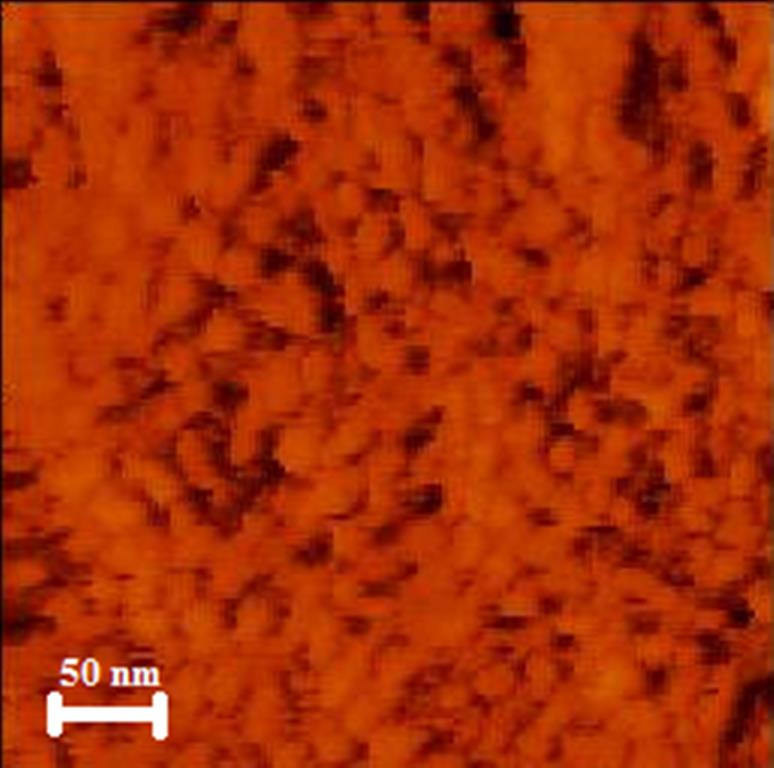}}
\caption{\label{fig:bob2} Scanning Tunneling Microscopy (STM) image of the Sb/Co/Sb (2.5 nm/0.5 nm/2.5 nm) sample after aging for 60 min. at 400K. \smallskip}
\end{figure}

\begin{table}[t]

 \begin{tabular}{| l | l | l | l |}
\hline
Measurement & Diameter (nm) & Perim (nm) \\ \hline

Mean &~~~~ 18.3  &~~~~ 57.6 \\ \hline

SD       &~~~~ 12.3  &~~~~ 45.4 \\ \hline

Min    &~~~~ 9.0  &~~~~ 25.4 \\ \hline

Max   &~~~~ 58.5  &~~~~ 254 \\ \hline

\end{tabular}
\\
\caption{ Geometric parameters for Before Aging Co nanoparticles found in a Sb/Co/Sb (2.5 nm/0.5 nm/2.5 nm) sandwich.}
\label{tab:multicol}
\end{table}

\begin{table}[t]

 \begin{tabular}{| l | l | l | l |}
\hline
Measurement & Diameter (nm) & Perim (nm)\\ \hline

Mean &~~~~ 19.2 &~~~~ 60.4\\ \hline

SD       &~~~~ 65.1 &~~~~ 677\\ \hline

Min    &~~~~ 9.0 &~~~~ 25.4\\ \hline

Max   &~~~~ 187 &~~~~ 673 \\ \hline

\end{tabular}
\\
\caption{ Geometric parameters for After Aging Co nanoparticles found in a Sb/Co/Sb (2.5 nm/0.5 nm/2.5 nm) sandwich.}
\label{tab:multicol}
\end{table}

\begin{figure} [t] 
\centering 
  \textbf{Sb/Co/Sb (2.5 nm/1.0 nm/2.5 nm)}\par\medskip
  \textbf{Before Aging}\par\medskip
\includegraphics[width=3.15in,height=2.8in,keepaspectratio]{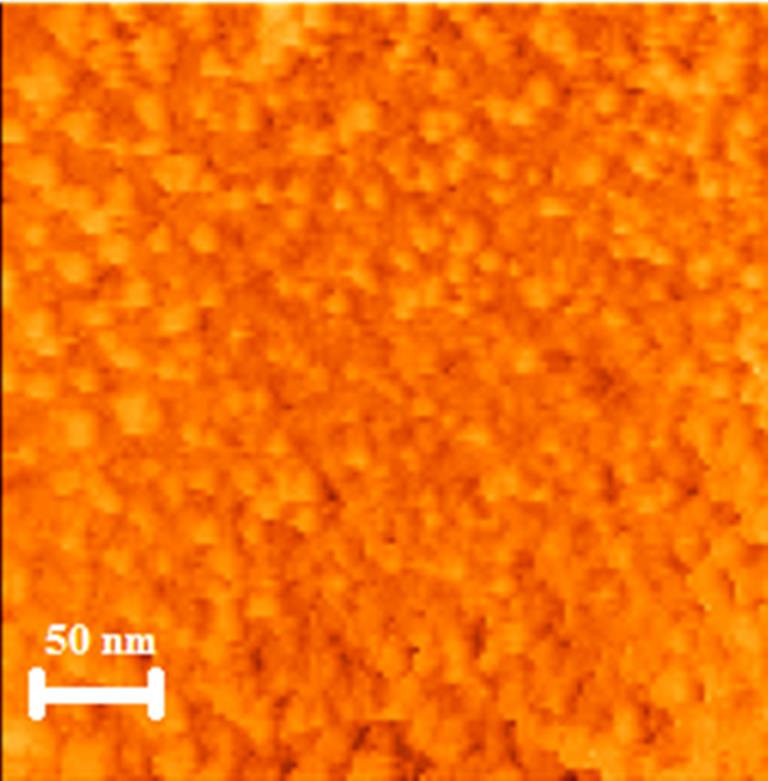}
\caption{\label{fig:bob1} Scanning Tunneling Microscopy (STM) image of the Sb/Co/Sb (2.5 nm/1.0 nm/2.5 nm) sample before aging.}

\end{figure}

\begin{figure}[t] 
\centering 
  \textbf{Sb/Co/Sb (2.5 nm/1.0 nm/2.5 nm)}\par\medskip
 \textbf{After Aging-a}\par\medskip
 \includegraphics[width=3.15in,height=2.8in,keepaspectratio]{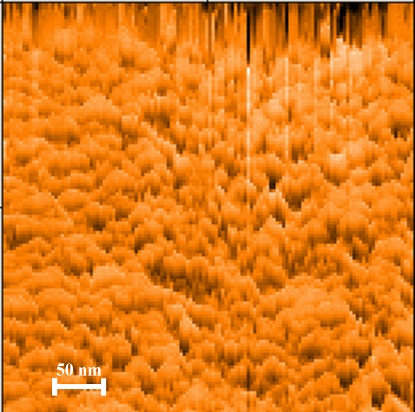}
\caption{\label{fig:bob2} Scanning Tunneling Microscopy (STM) image of the Sb/Co/Sb (2.5 nm/1.0 nm/2.5 nm) sample after aging for 60 min. at 400K. }

\end{figure}

\begin{table}[t]
\begin{tabular}{| l | l | l | l |}
\hline
Measurement & Diameter (nm) & Perim (nm)\\ \hline

Mean &~~~~ 21.9 &~~~~ 71.1\\ \hline

SD       &~~~~ 15.9 &~~~~ 61.6\\ \hline

Min    &~~~~ 9.0 &~~~~ 25.4 \\ \hline

Max   &~~~~ 85.4 &~~~~ 343 \\ \hline

\end{tabular}
\\
\caption{ Geometric parameters for Before Aging Co nanoparticles found in a Sb/Co/Sb (2.5 nm/1.0 nm/2.5 nm) sandwich.}
\label{tab:multicol}
\end{table}

\begin{table}

 \begin{tabular}{| l | l | l | l |}
\hline
Measurement & Diameter (nm) & Perim (nm)\\ \hline

Mean &~~~~ 22 &~~~~ 208\\ \hline

SD       &~~~~ 46 &~~~~ 819\\ \hline

Min    &~~~~ 9.0 &~~~~ 25.4 \\ \hline

Max   &~~~~ 96.9 &~~~~ 3048 \\ \hline

\end{tabular}
\\
\caption{ Geometric parameters for After Aging Co nanoparticles found in a Sb/Co/Sb (2.5 nm/1.0 nm/2.5 nm) sandwich.}
\label{tab:multicol}
\end{table}

\begin{figure}[b] 
\centering 
  \textbf{Sb/Co/Sb (2.5 nm/1.0 nm/2.5 nm)}\par\medskip
 \textbf{After Aging-b}\par\medskip

  \includegraphics[width=3.15in,height=2.8in,keepaspectratio]{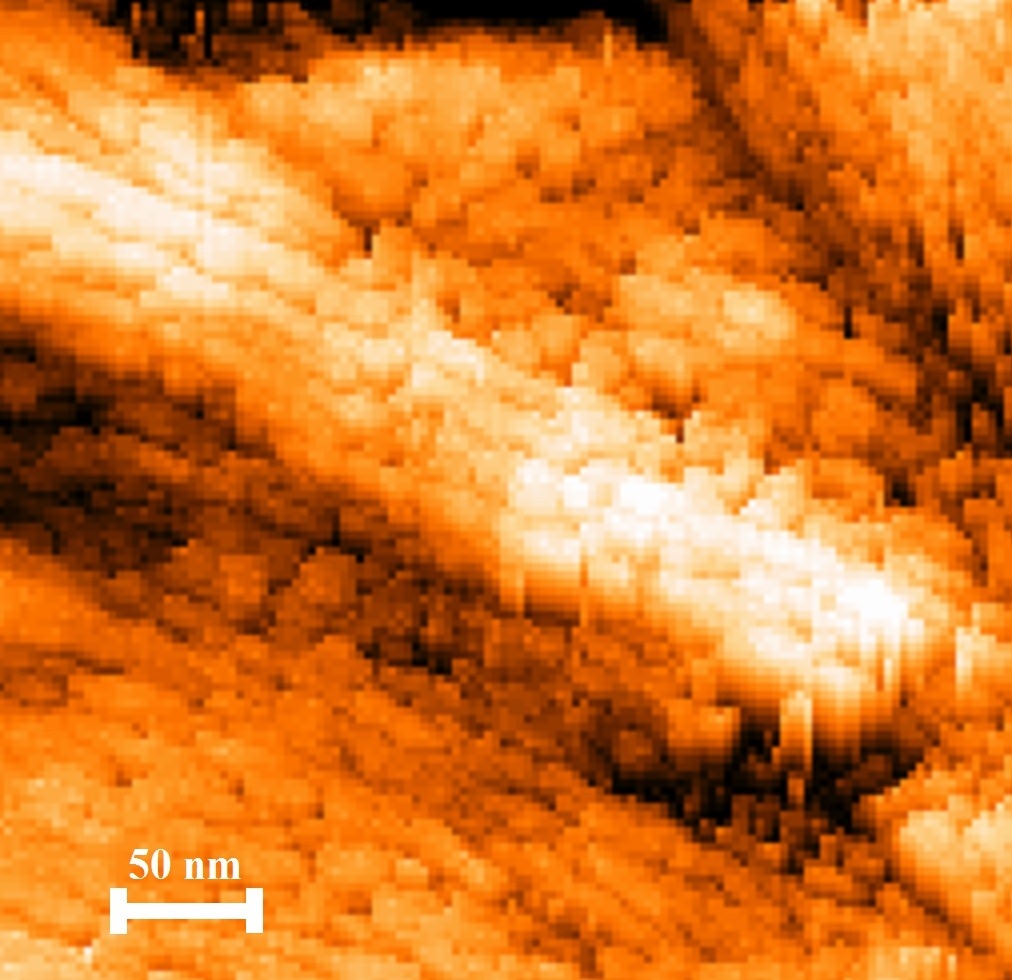}

\caption{\label{fig:bob2} Scanning Tunneling Microscopy (STM) image of the Sb/Co/Sb (2.5 nm/1.0 nm/ 2.5 nm) sample after aging for 60 min. at 400K. }
\end{figure}

\subsection{0.5 nm Co thickness}

Fig.~4 displays an STM image of the sample Sb/Co/Sb (2.5 nm/0.5 nm/2.5 nm) before aging. The figure clearly displays nanoscale particles that are fairly evenly distributed over the image, which is consistent with previous STM measurements \cite{Kenning11}. For Fig.~4, the threshold detection method observed 96 nanoparticles. Geometrical characteristics are presented in  Table II which displays the statistical data for the nanoparticle structure. The mean diameter of the particles was 18.3 nm. The ratio 
$r=perimeter/diameter = 3.15$, indicates that the particles are on average  circular in the plane.

Fig.~5 displays an STM image of the same sample (Fig.~4) after aging for 60 minutes at 400 K. For Fig.~5, the threshold detection method observed 46 nanoparticles. Geometrical characteristics are presented in  Table III which displays the statistical data for the nanoparticle structure.  The mean diameter before aging was 18.3 nm and after aging increased  about 5$\%$ to  19.2 nm. The standard deviation of the mean diameter and the perimeter increased markedly, indicating particles coalescing and formation of a few larger particles. This is confirmed by both the values of the maximum diameter and maximum perimeter. The ratio 
$r=perimeter/diameter = 3.15$ indicates that the detected particles are still quite circular after aging.

\begin{figure}[t] 
\centering 
  \textbf{Sb/Co/Sb (2.5 nm/5.0 nm/2.5 nm)}\par\medskip
  \textbf{Before Aging}\par\medskip
\centerline{\includegraphics[width=3.15in,height=2.8in,keepaspectratio]{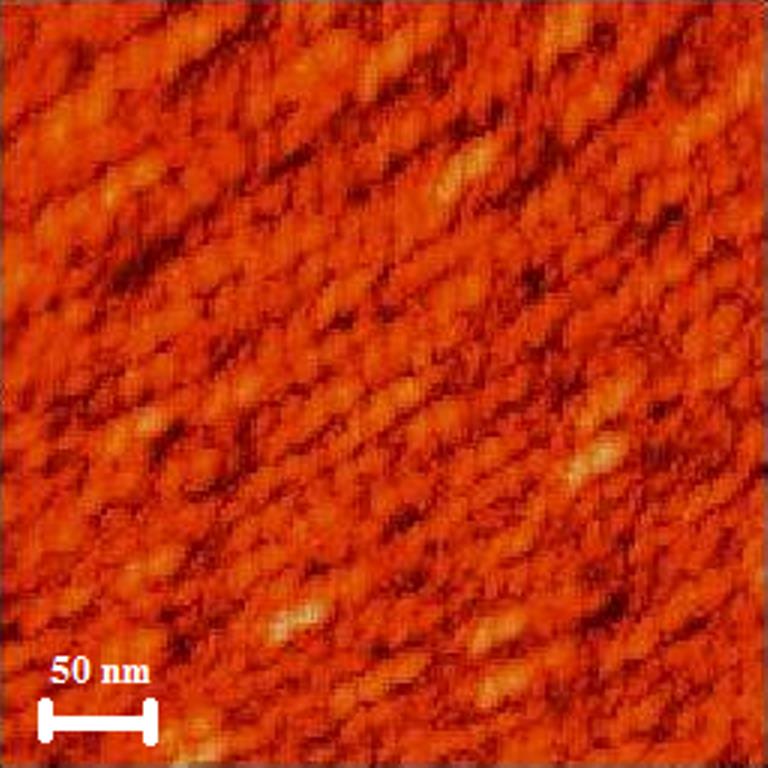}}
\caption{\label{fig:bob1} Scanning Tunneling Microscopy (STM) image of the Sb/Co/Sb (2.5 nm/5.0 nm/2.5 nm) sample before aging. \smallskip }
\end{figure}

\begin{table}[t]

 \begin{tabular}{| l | l | l | l |}
\hline
Measurement & Diameter (nm) & Perim (nm)\\ \hline

Mean &~~~~ 19.3 &~~~~ 62.3\\ \hline

SD       &~~~~ 17.7 &~~~~ 69.7\\ \hline

Min    &~~~~ 9.0 &~~~~ 25.4 \\ \hline

Max   &~~~~ 108 &~~~~ 419 \\ \hline

\end{tabular}
\\
\caption{ Geometric parameters for Before Aging Co nanoparticles found in a Sb/Co/Sb (2.5 nm/5.0 nm/2.5 nm) sandwich.}
\label{tab:multicol2}
\end{table}

\begin{figure}[t]
\centering 
  \textbf{Sb/Co/Sb (2.5 nm/5.0 nm/2.5 nm)}\par\medskip
 \textbf{After Aging}\par\medskip
 \includegraphics[width=3.15in,height=2.8in,keepaspectratio]{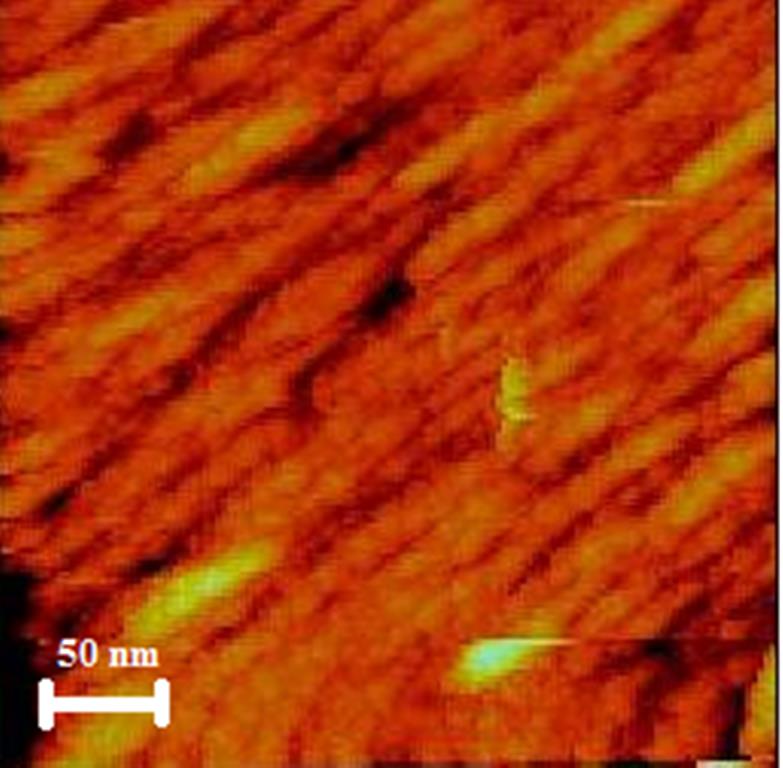}
\caption{\label{fig:bob2} Scanning Tunneling Microscopy (STM) image of the Sb/Co/Sb (2.5 nm/5.0 nm/2.5 nm) sample after aging for 60 min. at 400K. }

\end{figure}

\begin{table}[t]

 \begin{tabular}{| l | l | l | l |}
\hline
Measurement & Diameter (nm) & Perim (nm)\\ \hline

Mean &~~~~ 20.2 &~~~~ 61.3\\ \hline

SD       &~~~~ 18.2 &~~~~ 63.0\\ \hline

Min    &~~~~ 9.00 &~~~~ 25.4 \\ \hline

Max   &~~~~ 106 &~~~~ 368 \\ \hline

\end{tabular}
\\
\caption{Geometric parameters for After Aging Co nanoparticles found in a Sb/Co/Sb (2.5 nm/5.0 nm/2.5 nm) sandwich.}
\label{tab:multicol3}
\end{table}

In order to test the SPIP software, we made further measurements on the 0.5 nm pre-aged sample, displayed in Fig.~4, using the length tool located in the Nanosurf EasyScan E-STM scan panel. These mesurements were to provide an independent verification of the nanoparticle diameters. We also used the  Nanosurf profile analysis to measure the height of these particles. The measurements were made for approximately 43 particles. Our measurements for individual particles indicated a mean height of 6.0 nm. The mean diameter measured was 20.8 nm, which is about 13$\%$ greater than that obtained from the SPIP software. Values are reasonably close, giving us confidence in the SPIP analysis.

\subsection{1.0 nm Co thickness}

Fig.~6 displays an STM image of Sb/Co/Sb (2.5 nm/1.0 nm/2.5 nm) before aging. For Fig.~6, the threshold detection method observed 147 nanoparticles. Table IV  displays the statistical data for the nanoparticle structure. The mean diameter of the particles was 21.9 nm and the ratio 
$r=perimeter/diameter = 3.24$ indicates the particles are a little less circular in the plane than the Co (0.5 nm) samples.

Fig.~7 displays an STM image of the same sample (Fig.~6) after aging for 60 minutes at 400 K.  For Fig.~7, the threshold detection method observed 58 nanoparticles.  Table V  displays the statistical data for the nanoparticle structure.  The mean diameter after annealing was 22 nm. 
The standard deviation of the mean diameter and perimeter increased markedly, indicating particles coalescing with the formation of a few larger particles.  The ratio $r=perimeter/diameter = 9.45$ indicates that after aging the particles are no longer circular in the plane but have formed elongated structures. The r value plus the large SD, coupled with the extreme variation in particle perimeter values, suggest that the diameter calculation from the area (assumes that the particle is circular) is not valid for this sample.

In Fig.~8 we display a second STM image of the same sample (Fig.~6) after aging for 60 minutes at 400 K. In this image it is observed that the elevated structure in the middle of the image is made up of chains of nanoparticles that have  fused together in nanowire like structures. This is similar to what we observed in \cite{Kenning11}. It also appears that these nanowires may have fused to each other laterally producing a large coupled structure. The implications of the formation of large coupled structures will be discussed presently.

\subsection{5.0 nm Co thickness}

Fig.~9 displays an STM image of  Sb/Co/Sb (2.5 nm/5.0 nm/2.5 nm) before aging.  For Fig.~9, the detection method observed 113 nanoparticles. Geometrical characteristics are presented in Table VI which displays the statistical data for the nanoparticle structure.  The mean diameter of the particles was 19.3 nm and the ratio 
$r=perimeter/diameter = 3.22$ indicates the particles are reasonably circular in the plane. The particles appear to form in line segments, implying long range structural and possibly magnetic correlations.

Fig.~10 displays an STM image of the same sample (Fig.~9) after aging for 60 minutes at 400 K. For Fig.~10, the threshold detection method observed 46 nanoparticles. Geometrical characteristics are presented in Table VII which displays the statistical data for the nanoparticle structure.   The mean diameter before aging was 19.3 nm and after aging increased  about 5$\%$ to  20.2 nm. The ratio 
$r=perimeter/diameter = 3.03$ indicates the particles are reasonably circular in the plane. The particles arranged in the line segments (noted in the before aging sample) appear to flow together although the nanoparticles can still be discerned. It also appears that the line segments have a tendancy to clump together.

\section{Discussion}

We set out to determine if the variation of Co layer thickness in Sb/Co multilayer sandwiches affected the time-temperature characteristics of the resistance decay. We found that between Co layer thicknesses of 1.0 nm and 5.0 nm the characteristic timescale $\tau$, associated with the decay, increased by almost 100 times. This ability to vary the time-temperature characteristics offers an excellent opportunity for a time-temperature sensor tuned to the time-temperature characteristics of perishable items. The large variation enforces the need for accurate deposition parameters. For example, a variation of a monolayer of Co will cause an approximately eightfold variation in the characteristic timescale. To these results, we add a few observations from a concurrent study, where we have analyzed Sb/Co/Sb (2.5 nm/1.0 nm/2.5 nm) samples with Magnetic Force Microscopy (MFM), Atomic Force Microscopy (AFM) and Transmission Electron Microscopy (TEM). \cite{Kenning15} MFM measurements indicate that the  nanoparticles are magnetic and  single domain. AFM results indicate a before aging mean height of 7.8 nm and an after aging mean height of 9.0 nm. TEM measurements strongly indicate  the formation of a central core nanoparticle,  with a mean diameter before aging of 6.5 nm and an after aging mean core diameter of 8.0 nm. TEM also indicates that while Co nanoparticles form, much of the Co remains in the Co layer between the nanoparticles. This Co is observed to move and flow upon heating the sample (i.e. aging the sample). TEM also shows that some of the nanoparticles completely dissolve during aging, and flow in the Co layer.

To gain a further understanding of the effect of variations of the Co layer thickness, we performed STM scans of three different Co layer thicknesses (0.5 nm, 1.0 nm, and 5.0 nm) both before and after aging the sample.  The first thing that can be discerned, is that nano-sized structures are present. In all cases, nanoparticles  $\approx$20 nm in diameter are observed, covering over 80$\%$ of the surface. While these particles often appear very close to each other, in most cases  a thin dividing line between the particles can be discerned.  In all samples, the  particle diameter increased about 5$\%$ after aging and the number of particles per square decreased.   The particle diameters, as determined from the analysis of the STM images, are approximately three times larger than the core nanoparticle size determined from TEM imaging.  

A further issue to be addressed in understanding STM images is the sharp contrast between the density of states in Sb and the density of states in Co. The tunneling current modeled as a metal/insulating barrier/metal system is given by: 
\begin{eqnarray}
\begin{split}
I(V) =  \frac {4 \pi e}{\hbar} \int_{0}^{eV} [f(E_f -eV + \epsilon) - f(E_f + \epsilon)] \\
\times \rho_s (E_f - eV + \epsilon) \rho_T (E_f + \epsilon)|M|^2 d \epsilon 
\end{split}
\end{eqnarray}

where $f(E_f -eV + \epsilon)$ is the Fermi function and $\rho_s (E_f - eV + \epsilon)$ is the density of states in the sample while $f(E_f + \epsilon)$ and $\rho_T (E_f + \epsilon$ ) are the Fermi function and density of states of the tip. \cite{Suderow14}  The Fermi energy in Sb is near the bottom of the conduction band, or the top of the valence band, providing very few accessible states  for the tunneling current to populate.  Conversely, Co is a metal with a Fermi energy located in the conduction band, where there are large numbers of accessible states for the tunneling current to populate. The large contrast between the Sb and Co density of states is likely the reason we can observe the Co through the 2.5 nm Sb overlayer. Using XSTM, Kawasaki et al. \cite{Kawasaki11} looked at semimetal nanoparticles (ErAs) buried within a semiconductor (GaAs) and  observed signal of the particle at distances of a  1-2 nm from the nanoparticle. They attributed this effect  to electronic changes in the surrounding GaAs induced from the nanoparticle.   One of these effects, which may be important in the Co/Sb material, is band bending. Spin polarization of Sb at a NiMnSb/Sb interface out to $\approx$1 nm has been observed and described. \cite{Komesu00} We note that in a Co/Sb multilayer, with a 2.5 nm Sb layer sandwiched between two magnetic layers, the Sb might experience significant spin polarization.  
Unfortunately, the large contrast also makes accurate particle height measurements difficult. In Fig.~7 we observe a raised stucture which appears to be composed of a bundle of nanowires. We believe that this structure appears raised as the Co in the structure, not only has a much larger density of states at the Fermi energy than the Sb, but the density of states is also large compared to an individual nanoparticle. It is well known that the  density of states of a metal are strongly size dependent. \cite{Halperin86} As the nanowires in the structure coalesce, the size of the structure increases causing an increase in the Co density of states, thereby enhancing the contrast with the Sb.

We believe that once the Co is deposited on the Sb at 50$^oC$, the Co atoms begin to migrate towards a local nucleation point upon which a nanoparticle forms. After the Co is deposited we waited  300 sec. to allow more time for the Co to migrate. We associate the preparation time $t_1$ with the deposition time plus the waiting time.  It is during $t_1$  that a metastable Co state is formed.  Within a given area dominated by a particular nanoparticle, the Co atoms  migrate toward the nanoparticle, pulling away from regions associated with other nanoparticles. This would explain the particle separation observed with the STM. This state is then embedded  with the deposition of a 2.5 nm  Sb layer, effectively ending the preparation time. To understand the ``before aging" STM nanoparticle diameters, we conjecture that  we are forming a type of proto nanoparticle with a central core and a diffuse outer region formed by Co atoms. The diffuse outer region of atoms helps form an extended density of states for a particular nanoparticle. This extended state shows up in the STM images as an extended nanoparticle $\approx$20 nm. TEM measurements during aging indicate that, in general, the nanoparticles expand and begin to dissolve and some dissolve completely. These atoms, along with  Co atoms associated with the extended nanoparticle state, can flow between  nanoparticles, coupling them electrically and magnetically. Coupling the density of states over large regions within the Co will also contribute to decreasing the resistance. It is likely that this is the main contribution to the decrease in the initial resistance between the 2.5 nm and 5.0 nm Co layer thickness samples with the Co channel shorting out the effect of the resistance decays observed in Fig.~1. We believe that the resistance decays observed in Fig.~1 are primarily due to a semiconductor to semimetal transition in the Sb, which occurs during aging. \cite{Kenning14} We consider it reasonable that  the large magnetic moments interact with the band structure of the Sb  separating the conduction and valence bands, thereby producing a semiconductor. 

The reduction in the magnetization during aging may follow a similar mechanism to that proposed by Bedanta et al. \cite{Bedanta07}.  They suggest that small paramagnetic particles between the nanoparticles in 
$CoFe/Al_2O_3$ produce an antiferromagnetic coupling. It is also possible that the diffusion of atoms  in the region between the nanoparticles rearrange the spin structure of the extended system of coupled nanoparticles, significantly reducing the magnetization, thereby causing the semiconductor to semimetal transition.

In conclusion, we have measured the DC resistance decay of Co/Sb multilayers as a function of Co layer thickness. We find that for a given temperature, a five fold change in the Co thickness produces a 100 fold difference in the characteristic decay time of the resistance. STM imaging before and after aging indicate that the Co has evolved into nano-sized particles with evidence of an extended density of states beyond the nanoparticle core.

\section{Acknowledgments}

The authors would like to thank Dr. Devki Talwar for helpful discussions.

\end{document}